\documentclass[12pt,fleqn]{article}

\topmargin - 0.8cm

\catcode`@=11
\@addtoreset{equation}{section}
\catcode`@=12
\mathchardef\SGamma="7100

\begin{document}

\title{\bf Quantum Cauchy problem in cosmology}
\author{~A.~O.~Barvinsky $^a$\thanks{Email address: barvin@td.lpi.ru}
 ~and  ~D.~V.~Nesterov $^b$\thanks{Email address: nesterov@td.lpi.ru}}
\date{}
\maketitle
\begin{center}
\hspace{-8mm}$^a${\em
Theory Department, Lebedev Physics Institute and Lebedev Research Center in
Physics, Leninsky Prospect 53,
Moscow 117924, Russia}\\
\hspace{-8mm}$^b${\em
Moscow State University, Physics Faculty,
Department of Theoretical Physics \\
Moscow 119899, Russia.}
\end{center}

%%%%%%%%%%%%%%%%%%%%%%%%%%%%%%%%%%%%%%%%%%%%%%%%%%%%%%%%%%%%%%%%%%%%%%%%%%

\begin{abstract}
We develop a general framework for effective equations of expectation values
in quantum cosmology and pose for them the quantum Cauchy problem with
no-boundary and tunneling wavefunctions. We apply this framework in the
model with a big negative non-minimal coupling, which incorporates a
recently proposed low energy (GUT scale) mechanism of the quantum origin
of the inflationary Universe and study the effects of the quantum inflaton
mode.
\end{abstract}

%%%%%%%%%%%%%%%%%%%%%%%%%%%%%%%%%%%%%%%%%%%%%%%%%%%%%%%%%%%%%%%%%%%%%%%%%%
%%%  The body of the document.

\section{Introduction}
\hspace{\parindent}
Efim Samoilovich Fradkin created a scientific school that can be
characterized by a peculiar style that combines great diversity of
fundamental physical problems and efficient methods
of their solution. He never worked in the area of quantum cosmology, but
the generality of his methods, either it is the functional approach to QFT,
the Euclidean QFT or the most advanced methods of gauge constrained theories,
not only apply in this field but actually become indispensable. The authors
of this paper had a happy opportunity to share a small piece of his scientific
wisdom and, in this work, would like to show how these three
methods pioneered by E.S.Fradkin fruitfully combine in quantum
cosmological context.

The power of his approach is based on the fact that
the consideration of the problem in question begins with the first
fundamental principles followed by a set of clear mathematical
transformations which unambiguously push the solution to its
logical extreme. This strategy straightforwardly leads to the final result
instead of unreliable hand waving and (very often erroneous) insight.
In context of quantum cosmology this strategy implies a concrete setting of
the Cauchy problem for the initial cosmological state, the unitary
evolution from which gives rise to the present-day semiclassical Universe
with its observable cosmological parameters -- Hubble constant $H$,
density parameter $\Omega$, anisotropy of microwave background,
etc. Here we show how this strategy can be realized for a particular
model of inflationary Universe.
%by successively incorporating the methods of
%Euclidean QFT, Euclidean effective action, quantum reduction for constrained
%systems and effective equations for expectation values of gravitational and
%matter fields.

\section{Quantum origin of the Universe as a low-energy phenomenon}
\hspace{\parindent}
The virtue of quantum cosmology is that it can give initial conditions for
inflation, which in their turn
determine main cosmological parameters of the observable Universe, including
the density parameter $\Omega$. The requirement of $\Omega>1$ in closed
cosmology gives the bound on the e-folding number $N\geq 60$ -- the
logarithmic expansion coefficient for the cosmological scale factor $a$
during the inflation stage with a Hubble constant $H=\dot a/a$.
In the chaotic inflation model $H=H(\varphi)$ is generated by the inflaton
$\varphi$ and, therefore, all the parameters can be found as functions of
initial $\varphi$. This quantity is subject to the quantum distribution
$\rho(\varphi)$ determined by the cosmological wavefunction. If this
distribution has a sharp probability peak at certain $\varphi=\varphi_I$,
then this value serves as the initial condition for inflation.

Two known quantum states that lead in the semiclassical
regime to the closed inflationary Universe are the no-boundary
\cite{HH} and tunneling \cite{tun} wavefunctions. They both
describe quantum nucleation of the Lorentzian quasi-DeSitter spacetime
from the Euclidean hemisphere -- the gravitational
instanton responsible for the classically forbidden state of the
gravitational field. Tree level wave functions are generally devoid of the
observationally justified probability peaks.
Beyond the tree level the distribution $\rho_{\rm NB,T}(\varphi)$
becomes the diagonal element of the reduced density matrix obtained by
tracing out all degrees of freedom but $\varphi$ \cite{norm,BarvU}
        \begin{eqnarray}
        \rho_{\rm NB,T}(\varphi)\sim\exp[\mp I(\varphi)-
        \mbox{\boldmath$\SGamma$}(\varphi)].     \label{1.4}
        \end{eqnarray}
Here the classical action is amended by the Euclidean effective action
$\mbox{\boldmath$\SGamma$}(\varphi)$ of all quantum fields that are
integrated out.

The model of \cite{qsi} capable of generating the probability peak
contains the graviton-inflaton sector with a big negative constant,
$-\xi=|\xi|\gg 1$, of non-minimal curvature coupling,
        \begin{equation}
        {\mbox{\boldmath $L$}}(g_{\mu\nu},\varphi)
        =g^{1/2}\left\{\Big(\frac{m_{P}^{2}}{16\pi}
        -\frac{1}{2}\xi\varphi^{2}\Big)R
        -\frac{1}{2}(\nabla\varphi)^{2}
        -\frac{1}{2}m^{2}\varphi^{2}
        -\frac{\lambda}{4}\varphi^{4}\right\},      \label{1.5}
        \end{equation}
and generic GUT sector of Higgs $\chi$, vector gauge $A_\mu$ and spinor
fields $\psi$ coupled to the inflaton via the interaction terms
$\lambda_\chi\chi^2\varphi^2$, $g_{A}^2A_{\mu}^2\varphi^2$,
$f_{\psi}\varphi\bar\psi\psi$ (non-derivative parts) with the coupling
constants $\lambda_\chi,g_A,f_\psi$. Its peak-like distribution function
can be approximated by the gaussian packet
        \begin{eqnarray}
        \rho_{\rm NB,\,T}(\varphi)\simeq
        \frac1{\Delta\sqrt{2\pi}}
        \exp\left[-(\varphi-\varphi_I)^2/2\Delta^2\right], \label{12}
        \end{eqnarray}
where the parameters of the peak -- mean value
$\varphi_I= m_{P}(8\pi|1+\delta|/|\xi|\mbox{\boldmath$A$})^{1/2}$ and
its quantum width $\Delta= (\varphi_I/\sqrt{12{\mbox{\boldmath $A$}}})
\sqrt{\lambda}/|\xi|$ are strongly suppressed by a small ratio
$\sqrt{\lambda}/|\xi|$ known from the COBE normalization for
$\Delta T/T\sim 10^{-5}$ \cite{COBE} (because the CMBR anisotropy
in this model is proportional to this ratio \cite{nonmin1}). Here
${\mbox{\boldmath $A$}} = 1/2\lambda
        \Big(\sum_{\chi} \lambda_{\chi}^{2}
        + 16 \sum_{A} g_{A}^{4} - 16
        \sum_{\psi} f_{\psi}^{4}\Big)$
is the universal combination of coupling constants above and
        \begin{eqnarray}
        \delta\equiv
        -8\pi\,|\xi|\,m^2/\lambda\,m_P^2.   \label{delta}
        \end{eqnarray}
For the no-boundary and tunneling states ($\pm$-signs respectively) the peak
exists for positive $\mbox{\boldmath$A$}$ and $\pm(1+\delta)<0$. The
classical equations of motion in the slow roll approximation,
        \begin{eqnarray}
        \ddot{\varphi}+3H(\varphi)\dot{\varphi}-F(\varphi)=0, \label{1.12}
        \end{eqnarray}
$H(\varphi)\simeq(\lambda/12|\xi|)^{1/2}\varphi$,
$F(\varphi)\simeq-\lambda m_P^2(1+\delta)\varphi/48\pi\xi^2$,
show that the inflaton decreases from its initial value,
$\dot\varphi\simeq F/H<0$, only for $1+\delta>0$, that is only for the
{\em tunneling} state. Only in this case
the duration of the inflationary epoch is finite with the e-folding number
\cite{efeq} $N=\int dt H\simeq 48\pi^2/\mbox{\boldmath $A$}$.
Comparison with $N\geq 60$ yields the bound $\mbox{\boldmath $A$}\sim 5.5$
which can be regarded as a
selection criterion for particle physics models \cite{qsi}. This conclusion
remains qualitatively true when taking into account the contribution of
the inhomogeneous quantum modes to the radiation current of theeffective
equations \cite{efeq}. This contribution and its dynamical effect
were obtained in \cite{efeq} by the method of the Euclidean effective action,
however, the quantum fluctuations of the inflaton field itself have not been
taken into account.

For the proponents of the no-boundary quantum states in a long debate on the
wavefunction discord \cite{VilVach,discorde} this
situation seems unacceptable. According to this result the
no-boundary proposal does not generate realistic inflationary scenario,
while the tunneling state does not satisfy important aesthetic criterion
-- the universal formulation of the initial conditions and dynamics
in one concept -- spacetime covariant path integral over
geometries.
Thus, one of the motivations of considering the quantum mechanical
sector of the inflaton mode is the hope that it can handle this difficulty.
In view of the smallness of $\Delta$ the quantum
fluctuations $\Delta\varphi\sim\Delta$ are expected to be negligible, but
those of their quantum momenta $\Delta p_\varphi\sim 1/\Delta$ blow up
for small $\Delta$. Therefore, apriori, it is hard to predict the overall
magnitude of the quantum rolling force and its sign due to
$\Delta\varphi(t)$. In what follows we carefully consider this problem.

\section{Effective equations: setting the problem}
\hspace{\parindent}
Effective equations for expectation values of quantum
fields, $\hat{g}(x)=\hat{\varphi}(x)$, $\hat{\chi}(x),
\hat{\psi}(x),\hat{A}_\mu(x),\hat{g}_{\mu\nu}(x),...$, in the
quantum state $|{\mbox{\boldmath $\Psi$}}\big>$,
        $g(x)=\big<{\mbox{\boldmath $\Psi$}}|
        \hat{g}(x)
        |{\mbox{\boldmath $\Psi$}}\big>$,
can be obtained by expanding the Heisenberg equations of motion for
$\hat{g}(x)=g(x)+\Delta\hat{g}(x)$ in powers of
$\Delta\hat{g}(x)$ and averaging them with respect to
$|{\mbox{\boldmath $\Psi$}}\big>$:
        \begin{eqnarray}
        \frac{\delta S[\,g\,]}{\delta g(x)}
        +J(x)=0.                              \label{2.3}
        \end{eqnarray}
Here $S[\,g\,]$ is the classical action and $J(x)$ is the radiation
current which accumulates all quantum corrections starting with the
one-loop contribution
        \begin{eqnarray}
        J(x)=\frac 12\int dy\,dz\,
        \frac{\delta^3 S[\,g\,]}
        {\delta g(x)\,\delta g(y)\,\delta g(z)}
        \,G(z,y)+...  =\frac 12\,
        \left<\,\left[\,\frac{\delta S[\hat g]}
        {\delta\hat g(x)}\,\right]_2 \right>+...,      \label{2.6}
           \label{2.4}
        \end{eqnarray}
and $G(z,y)=\big<{\mbox{\boldmath $\Psi$}}|\,\Delta\hat{g}(z)\,
\Delta\hat{g}(y)\,|{\mbox{\boldmath $\Psi$}}\big>$ is the Wightman function
of quantum disturbances in a given quantum state.

In closed cosmological model, the total metric and the scalar field (playing
the role of an inflaton) is usually decomposed in the spatially homogeneous
background and inhomogeneous perturbations
        $\, ds^2=-N^2(t)dt^2+a^2(t)\gamma_{ij}dx^i dx^j
        +h_{\mu\nu}(x)dx^\mu dx^\nu$,
        $\;\,\varphi(x)=\varphi(t)+\delta\varphi(x)$,
where $a(t)$ is the scale factor, $N(t)$ is the lapse function and
$\gamma_{ij}$ is the spatial metric of the 3-sphere of unit radius. Therefore,
the full set of fields $g(x)$ consists of the minisuperspace
spatially homogeneous variables $Q(t)$ and inhomogeneous fields $f(x)$
depending on spatial coordinates $x^i$={\bf x},
        $\;Q(t)=(a(t),\,\varphi(t),\,N(t))$,
        $\;f(x)=(\delta\varphi(t,\mbox{\bf x}),
        \,h_{\mu\nu}(t,\mbox{\bf x}),
        \,\chi(t,\mbox{\bf x}),\,\psi(t,\mbox{\bf x}),\,
        A_\mu(t,\mbox{\bf x}),...)$.
From the structure of the initial quantum state, that will be discussed later,
it follows that only minisuperspace variables have nonvanishing expectation
values $\big<\,\hat{Q}(t)\,\big>\neq 0,\,\,\,\,\big<\,\hat{f}(x)\,\big>=0$.
Therefore, the full set of effective equations reduces to the following three
equations in the minisuperspace sector
        \begin{eqnarray}
        &&\frac{\delta S\,[Q]}{\delta Q(t)}+J_Q(t)=0,
        \qquad
        J_Q=J_N,\;J_a,\;J_\varphi,                     \label{2.10}
        \end{eqnarray}
their quantum radiation currents $J_Q(t)$ containing the contribution of
quantum fluctuations of minisuperspace modes themselves and those of the
spatially inhomogeneous fields.

Splitting of the whole configuration space into minisuperspace and
inhomogeneous sectors reflects the choice of the collective
variables and their quantum states -- they turn out to be very
different for these two sectors of the theory. This results in different
calculational strategies for the corresponding quantum averages. Inhomogeneous
$f$-modes live in the Euclidean DeSitter invariant vacuum, so that
their calculation is strongly facilitated by the Euclidean effective action
method \cite{efeqmy}. For the quantum mechanical $\varphi$-mode the
situation is different.

In the tree-level approximation the wavefunction $\exp[\mp I(\varphi)/2]$
of $\varphi$ does not have good probability peaks and
is even unnormalizable. Therefore, the tree-level quantum average
$\big<\Delta Q\Delta Q\big>^{\rm tree}$ is
badly defined. Beyond the tree-level approximation the situation can be
improved, because the quantum average should now be defined with the aid
of the reduced density matrix $\big<\Delta Q\Delta Q\big>=
{\rm tr}\left[\Delta\hat Q\Delta\hat Q\,\hat\rho\right]$,
        \begin{eqnarray}
        \hat\rho\equiv\rho(\varphi,\varphi')=\int df
        \mbox{\boldmath$\Psi$}(\varphi,f)
        \mbox{\boldmath$\Psi$}^*(\varphi',f),         \label{2.25}
        \end{eqnarray}
which originates from tracing the $f$-variables out and includes loop
corrections. As shown in \cite{norm,decoh}, the diagonal
element of this density matrix -- the distribution function of $\varphi$,
$\;\rho(\varphi)=\rho(\varphi,\varphi)$,
is given in the approximation of a gaussian integral by the
effective action algorithm (\ref{1.4}) which can generate a sharp
probability peak (\ref{12}). With this modification the
quantum correlators become well defined, being expressed in terms of
$\big<\Delta\varphi\Delta\varphi\big>\sim\Delta^2<\infty$.

\section{Quantum Cauchy problem: tree level approximation}
\hspace{\parindent}
In the one-loop approximation the radiation current can be calculated on the
classical background -- the lowest order approximation for the mean field.
The initial conditions for this solution follow from the no-boundary and
tunneling wavefunctions. We show this for the model of
a minimally coupled inflaton field $\phi$ with a generic potential $V(\phi)$
(we reserve the notation $\phi$ as opposed to the notation $\varphi$ for the
non-minimal inflaton\footnote{This framework can be extended
to the non-minimal model by reparametrizing the latter to the
Einstein frame \cite{renorm}, and this will be done below.}). The
Wheeler-DeWitt equation for the cosmological wavefunction on 2-dimensional
minisuperspace, $\mbox{\boldmath$\Psi$}(\phi,a)$, has two semiclassical
solutions -- the so-called no-boundary and tunneling wavefunctions. In the
approximation of the inflationary slow roll (when the derivatives with
respect to $\phi$ are much smaller than the derivatives with respect to $a$)
they read \cite{VilVach}
        \begin{eqnarray}
        &&\mbox{\boldmath$\Psi$}_{NB}(\phi,a)=
        C_{NB}(a^2H^2(\phi)-1)^{-1/4}e^{-I(\phi)/2}
        \cos\left[S(a,\phi)+\frac\pi4\right],        \label{3.11}  \\
        &&\mbox{\boldmath$\Psi$}_{T}(\phi,a)=
        C_{T}(a^2H^2(\phi)-1)^{-1/4}
        e^{I(\phi)/2}\exp\left[iS(a,\phi)+\frac{i\pi}4\right]
\label{3.12}
        \end{eqnarray}
and describe the nucleation of the Lorentzian DeSitter spacetime with the
Hamilton-Jacobi function
$S(\phi,a)=-\pi m_P^2(a^2H^2(\phi)-1)^{3/2}/2H^2(\phi)$ from the
gravitational half-instanton with the action $I(\phi)/2=-\pi m_P^2/H^2(\phi)$.
This nucleation takes place at $a=1/H(\phi)$,
$H(\phi)\equiv(\kappa V(\phi)/3)^{1/2}$ being the effective Hubble constant,
driving the inflationary dynamics of the model, $\dot{a}/a\simeq H(\phi)$
($\kappa\equiv8\pi/m_P^2$ -- the gravitational constant). This domain forms
the one-dimensional curve in two-dimensional minisuperspace,
        \begin{eqnarray}
        \chi(\phi,a)=a-(3/\kappa V(\phi))^{1/2}=0,   \label{3.15}
        \end{eqnarray}
which can be identified with the physical subspace, $\chi(\phi,a)$ being
regarded as the corresponding gauge condition. If $\phi$ is chosen as a
physical coordinate then, according to the formalism of \cite{BKr}, the
physical wavefunction $\mbox{\boldmath$\Psi$}(\phi)$ follows from the
Dirac wavefunction $\mbox{\boldmath$\Psi$}(\phi,a)$ by the transformation
        $\mbox{\boldmath$\Psi$}(\phi)=
        J_\chi^{1/2}
        \mbox{\boldmath$\Psi$}(\phi,a)|_{\chi(\phi,a)=0}$,
where $J_\chi$ is the Faddeev-Popov determinant -- the Poisson bracket of
the gauge condition with the Hamiltonian constraint. Since
$J_\chi^{1/2}\sim(H^2a^2-1)^{1/4}$, it cancels the divergent preexponential
factors in (\ref{3.11})-(\ref{3.12}) and results in the initial physical
wavefunctions $\mbox{\boldmath$\Psi$}_{NB,T}(\phi)\sim\exp[\mp I(\phi)/2]$.
Beyond the tree level they get replaced by the density matrix (\ref{2.25})
which, for the model of the nonminimally coupled inflaton, even yields a
sharp probability peak and, thus, generates the expectation value of the
inflaton $\phi=\big<\hat\phi\big>$ -- the first initial condition of the above
type.

The second initial condition arises from the expectation
value of the {\em physical} momentum conjugated to $\phi$,
$\;p_\phi=\big<{\hat p}_\phi\big>$. In view of reality of the initial
density matrix (\ref{2.25}) this expectation value is vanishing,
        $\Big<\hat{p}_\phi\Big>=0$.
From the Hamiltonian reduction of the symplectic form in the gauge
$\chi(\phi,a)=0$ it follows that the physical momentum expresses in terms
of the original momenta, $\Pi_a da+\Pi_\phi d\phi=p_\phi d\phi$,
        $\;p_\phi=\Pi_\phi-\Pi_a \chi_\phi/\chi_a$,
        $\;\chi_\phi\equiv\partial_\phi\chi$,
        $\;\chi_a\equiv\partial_a\chi$.
Therefore, for $p_\phi=0$, $\;\Pi_\phi$ homogeneously expresses in terms of
$\Pi_a$ and, in view of the Hamiltonian constraint this implies that at
the initial Cauchy surface $\Pi_\phi=0$ and $\Pi_a=0$. Thus, the full set
of initial conditions for the classical background reads as
        $\phi=\big<\hat\phi\big>,\,\,\,
        a=1/H(\phi),\,\,\,\dot\phi=\dot a=0$.

\section{Cauchy problem for linearized quantum fluctuations}
\hspace{\parindent}
In the Hamiltonian reduction to the physical sector for cosmological
perturbations \cite{GMTS}, scalar perturbations
of metric and inflaton fields (which only contribute to the homogeneous
sector) are defined by the ansatz:
        $ds^2_{\rm total}=a^2(\eta)\left[-(1+2A)d\eta^2
        +(1-2\psi)\gamma_{ij}dx^idx^j\right]$,
        $\;\phi_{\rm total}
        =\phi+\delta\phi$.
Their quadratic action reduces to the
functional of only two invariant (with respect to linearized gauge
transformations) combinations of canonical coordinates
$(\psi,\delta\phi)$ and their momenta $(\Pi_\psi,\Pi_{\delta\phi})$,
$\;\mbox{\boldmath$\Psi$}=\psi+{\cal H}\delta\phi/\phi'$,
$\;\mbox{\boldmath$\Pi$}_\psi=\Pi_\psi-2a^2\sqrt{\gamma}
D\delta\phi/\kappa\phi'$, where prime denotes derivatives with
respect to the conformal time $\eta$, ${\cal H}=a'/a$ is the
conformal time Hubble constant and $D=\gamma^{ij}\nabla_i\nabla_j+3$.
With the choice of physical phase space variables,
        $\mbox{\boldmath$q$}=
        2a\mbox{\boldmath$\Psi$}/\kappa\phi'
        +{\cal H}D^{-1}
        \mbox{\boldmath$\Pi$}_\psi/a\phi'\sqrt{\gamma}$,
        $\mbox{\boldmath$p$}=-\phi'a\sqrt{\gamma}
        D\mbox{\boldmath$\Psi$}/{2\cal H}
        +\kappa\phi'\mbox{\boldmath$\Pi$}_\psi/4a$,
the action acquires the final Lagrangian form
        \begin{eqnarray}
        S[\mbox{\boldmath$q$}]\Big|_2
        =\frac12\int d\eta\,\sqrt{\gamma}(-D\mbox{\boldmath$q$})
        \Big[-d^2/d\eta^2+\phi'(1/\phi')''
        +\kappa\phi'^2/2+D\Big]\mbox{\boldmath$q$}.   \label{4.16}
        \end{eqnarray}
The field $\mbox{\boldmath$q$}$ is well known from the theory of cosmological
perturbations as the Bardeen invariant \cite{GMTS}\footnote{Note that the
operator $D$ is positive definite for the spatially homogeneous mode, $D=+3$.
Thus, this is a ghost variable signifying the classical instability.
This instability at the linear level is the
manifestation of inflation which is a huge instability phenomenon
incorporating the runaway modes. In contrast with the
S-matrix theory this instability should not be regarded as an irrecoverable flaw
of the theory, because we know a nonlinear mechanism that provides
a graceful exit from the inflation stage. In particular, no special measures
like introducing the indefinite metric should be undertaken to eradicate
this phenomenon. Homogeneous fluctuations of the inflaton field do not have
a particle nature and one should not take care of guaranteeing the energy
positivity of their excitations. Therefore, this mode can and should be
quantized in the coordinate representation with positive metric in the
Hilbert space.}.

In the Newton gauge, widely used in the theory of cosmological perturbations,
these perturbations read in terms of the Bardeen invariant
$\mbox{\boldmath$q$}$ as \cite{GMTS}
        $A=\psi$,
        $\psi=\kappa\phi'\mbox{\boldmath$q$}/2a$,
        $\delta\phi=
        (\phi'\mbox{\boldmath$q$})'/a\phi'$.
Similar relations in the minisuperspace gauge (\ref{3.15})
can be obtained by expressing $\delta a=-a\psi$ in terms of $\delta\phi$,
$\psi=(\chi_\phi/a\chi_a)\delta\phi$, and finding the expression
for the physical momentum $p_{\delta\phi}$ conjugated to $\delta\phi$,
$\Pi_\psi\psi'+\Pi_{\delta\phi}\delta\phi'=p_{\delta\phi}\delta\phi'
+...$, $p_{\delta\phi}=\Pi_{\delta\phi}+\Pi_\psi\chi_\phi/a\chi_a$.
Then, the linearized Hamiltonian constraint gives
$\Pi_\psi$ and $\Pi_{\delta\phi}$ as functions of
$(\delta\phi,p_{\delta\phi})$, and one directly proceeds to the
transformation relating $(\mbox{\boldmath$q$},\mbox{\boldmath$p$})$
to $(\delta\phi,p_{\delta\phi})$. For $\eta\to 0$ this transformation turns
out to be singular (see \cite{efeqmy} for details). This singularity is,
however, an artifact of the definition of the invariant variables non-analytic
at $\phi'\to 0$. To see this, we decompose the general solution for
$\mbox{\boldmath$q$}(\eta)$ in the sum of two linearly independent solutions
of the equation of motion for the action (\ref{4.16}),
        $\mbox{\boldmath$q$}(\eta)
        =c_+\mbox{\boldmath$q$}_+(\eta)+
        c_-\mbox{\boldmath$q$}_-(\eta)$,
one of them having a singular behaviour at $\eta=0$,
$\mbox{\boldmath$q$}_+(\eta)=\eta^3/\phi'\left(1+O(\eta^2)\right)$,
$\mbox{\boldmath$q$}_-(\eta)=1/\phi'\left(1-3\eta^2/2+O(\eta^3)\right)$.
Then we show that the coefficients
$c_{\pm}$  have a regular solution in terms of the initial conditions for
physical variables $(\delta\phi(0),p_{\delta\phi}(0))$ \cite{efeqmy},
        $\;c_+=-V_\phi p_{\delta\phi}(0)/9H\sqrt\gamma$,
        $\;c_-=V_\phi\delta\phi(0)/3H^{3/2}$.
This relation will be used throughout the rest of the paper to
express the Heisenberg operators
$\Delta\hat Q_{\rm phys}(\eta)$ and $\Delta\hat Q(\eta)$ in terms of
the Schrodinger ones,
$\delta\hat\phi(0)=\delta\phi,\;
\hat p_{\delta\phi}(0)=\partial/i\partial(\delta\phi)$, and then find the
quantum averages of their bilinear combinations in the
$\delta\phi$-representation of the initial density matrix
(see eq.(\ref{12})
with $\delta\phi=\phi-\big<\phi\big>$).

\section{Non-minimal model}
\hspace{\parindent}
The action of the model (\ref{1.5}) has a generic form
        \begin{eqnarray}
        S[g_{\mu\nu},\varphi]=\int d^4x\, g^{1/2}\left\{-V(\varphi)
        +U(\varphi)R
        -\frac12 G(\varphi)(\nabla\varphi)^2\right\},       \label{5.1}
        \end{eqnarray}
where the coefficient functions can be read off (\ref{1.5}). In the presence
of spatial densities of one-loop radiation currents
        $j_N \equiv J_N/a^3\sqrt\gamma$,
        $\;j_\varphi\equiv J_\varphi/Na^3\sqrt\gamma$,
the effective equation for the inflaton field reads
\cite{efeqmy}
        \begin{eqnarray}
        &&\ddot\varphi+\Big(3\,\frac{\dot a}a
        -\frac{a}{2\dot a}\,
        U_\varphi\,j_\varphi\Big)\dot\varphi
        -F(\varphi,a,\dot\varphi)=0,                    \label{5.36}\\
        &&F(\varphi,a,\dot\varphi)=
        \frac{2VU_\varphi\!-\!UV_\varphi}
        {U+3U_\varphi^2}\!-\!\frac{\dot\varphi^2}2\frac d{d\varphi}
        \ln(U+3U_\varphi^2)\!
        +\!F_{\rm loop}(\varphi,a,\dot\varphi,\dot a),   \label{5.36a}\\
        &&F_{\rm loop}(\varphi,a,\dot\varphi,\dot a)
        =\frac1{U+3U_\varphi^2}\left(Uj_\varphi
        -2U_\varphi\,j_N
        -\frac a{2\dot a}\frac{d\,j_N}{dt}\right).    \label{5.37}
        \end{eqnarray}
It contains quantum contributions to the friction term and the rolling
force, while the first two terms in (\ref{5.36a}) represent the classical
part. As regards the quantum part, one-loop radiation
currents split into the contributions of the
quantum mechanical sector and the sector of spatially inhomogeneous
modes, $j_{\rm 1-loop}=j^q+j^f$. The $f$-part of the current can be
generated by the effective action, which implies the replacement of the
classical coefficient functions $V(\varphi),U(\varphi),G(\varphi)$, by their
effective counterparts
and truncation of the (generally infinite) series to the first three terms.
This truncation is based on two assumptions -- the smallness of inflaton
derivatives due to the slow roll regime and smallness of $R/m^2_{\rm part}$
-- the curvature to particle mass squared ratio \cite{efeq}.
Thus, with this approximation, the effective equations of motion in our
non-minimal model take the form of (\ref{5.36}) with
$V_{\rm eff}(\varphi),U_{\rm eff}(\varphi),G_{\rm eff}(\varphi)$ replacing
$V(\varphi),U(\varphi),G(\varphi)$ and the radiation currents $j_N,j_\varphi$
saturated by the contribution of the quantum mechanical mode,
$j_N^q,j_\varphi^q$. The resulting rolling force in the leading order
of the slow roll expansion becomes the sum of the force induced by the
effective action, $F^{\rm eff}$, and the quantum mechanical force, $F^q$,
$F=F^{\rm eff}+F^q$,
        \begin{eqnarray}
        F^{\rm eff}=\frac{2V^{\rm eff}U_\varphi^{\rm eff}
        -U^{\rm eff}V_\varphi^{\rm eff}}
        {G^{\rm eff}U^{\rm eff}+3(U^{\rm eff}_\varphi)^2},\qquad
        F^q=\frac1{U+3U_\varphi^2}\left(Uj^q_\varphi
        -2U_\varphi\,j_N^q
        -\frac a{2\dot a}\frac{d\,j_N^q}{dt}\right).    \label{5.37b}
        \end{eqnarray}

\section{Radiation currents and their effect on inflationary dynamics}
\hspace{\parindent}
The role of $F^{\rm eff}$ in the inflationary
evolution has been studied in \cite{efeq}. For the no-boundary and tunneling
states in the one-loop approximation it equals
        \begin{eqnarray}
        F^{\rm eff}_{NB,T}=-\frac{\lambda m_P^2
        (1+\delta)}{48\pi\xi^2}\,\varphi
        \left(1\mp\frac{\varphi^2}{\varphi^2_I}\right)
        +O(1/|\xi|^3)                                \label{5.37g}
        \end{eqnarray}
and leads to different conclusions. The no-boundary peak is realized for
$1+\delta<0$, therefore the point $\varphi_I$ is an attractor -- quantum
terms in $F^{\rm eff}$ lock the inflaton at its constant initial
value and give rise to infinitely long inflationary scenario with exactly
DeSitter spacetime. In the tunneling case, the probability peak
exists in the opposite range
of the parameter (\ref{delta}), $\delta>-1$, and the rolling force
has the quantum term which initially doubles the negative classical part.
Therefore, the inflaton decreases under the influence of
this force, and the tunneling state generates a finite inflation stage
with the estimated e-folding number
$N\simeq 48\pi^2\ln 2/\mbox{\boldmath$A$}$ leading to the estimate on
$\mbox{\boldmath$A$}$, $\mbox{\boldmath$A$}\leq 5.5$. As we shall now see,
the quantum mechanical force $F^q$ does not qualitatively change these
predictions.

To obtain radiation currents in (\ref{5.37b}) we expand the first
order variations of (\ref{5.1})
to the second order in perturbations $(A,\psi,\delta\varphi)$
and make their quantum averaging with respect to the initial state. For
this purpose we, first, need the initial reduced density
matrix of the inflaton field $\rho(\varphi,\varphi')$ in the non-minimal
model and, second, the expressions for the Heisenberg operators
$\Delta\hat Q(\eta)$ in terms of quantum initial data,
$\delta\hat\varphi=\delta\varphi$,
$\;\hat p_{\delta\varphi}=\partial/i\partial(\delta\varphi)$. In \cite{decoh}
it was shown that for the model with a big $|\xi|$ the initial
density matrix describes practically pure quantum state and expresses in
terms of the distribution function
$ \rho(\varphi,\varphi')\simeq
\rho^{1/2}(\varphi)\rho^{1/2}(\varphi')$, $\;|\xi|\gg 1$.
The effective pure quantum state,
$\mbox{\boldmath$\Psi$}_{NB,T}(\varphi)\simeq\rho_{NB,T}^{1/2}(\varphi)$,
in the vicinity of the probability maximum, which is located at $\varphi_I$,
can, thus, be approximated by the gaussian packet of small quantum width
$\Delta$ -- the square root of (\ref{12}).

The operators of quantum disturbances in the $\delta\varphi$-representation
can be found by using the results of Sects.4-5 obtained for the minimal
model because it can be regarded as the Einstein frame for the
non-minimal one. The Einstein frame for (\ref{5.1}) with $\bar U=m_P^2/16\pi$
and $\bar G=1$ arises by a special conformal transformation and
reparameterization of the inflaton field \cite{renorm}:
$(g_{\mu\nu},\varphi)\rightarrow(\bar g_{\mu\nu},\bar\varphi)$.
We denote the objects in the Einstein frame by bars and identify them with
those of the minimal model considered in Sects.4-5. In this way we
reduce all the calculations, Cauchy data setting, gauge fixing, reduction
to the physical sector, etc. to those of the minimal model. This makes the
further calculations of radiation currents straightforward. Below
we separately consider the beginning of the inflation epoch, $t=0$, and late
stationary stage of inflation.

At the onset of inflation $t=0$ the radiation currents equal \cite{efeqmy}
        \begin{eqnarray}
        j_N^q(0)=\frac{\lambda^2\varphi_I^4}{384\pi^2|\xi|^2}
        \left(\frac1f-\frac13 f\right),  \qquad
        j_\varphi^q(0)=
        \frac{\lambda^2\varphi^3_I}{96\pi^2|\xi|^2}
        \left(\frac1f-\frac16 f\right),        \label{6.9}
        \end{eqnarray}
and $(a/\dot a)dj_N^q/dt(0)=0$, where $f=(\mbox{\boldmath $A$}/
16\pi^2)^2|\xi|/|1+\delta|$. These quantities are strongly suppressed as
compared to their classical values, by a very small
factor $\lambda/|\xi|^2\sim\Delta T^2/T^2\sim 10^{-10}$ related to the
CMBR anisotropy \cite{nonmin1}. Their sign crucially depends on the magnitude
of the parameter $f$, which in our model is likely to be very big,
$f\gg 1$ (in view of the estimate $N\geq 60$ on the e-folding number
and the value of $|\xi|\sim 10^4$ \cite{nonmin1}).
In this case, the terms proportional to $f\sim 1/\Delta^2$, generated by the
kinetic terms of the radiation currents, $\Big<\Delta Q'\Delta Q'\Big>$,
dominate and, in particular, lead to the positive energy density,
        $\varepsilon^q(0)=-j_N(0)\simeq
        m_P^4\lambda^2|1+\delta|/64\pi^2|\xi|^3\ll m_P^4$, and, as
one can check \cite{efeqmy}, negative pressure $p^q(0)=-\varepsilon^q(0)$
(DeSitter equation of state). Interestingly, the sign of the quantum
rolling force due to the homogeneous mode is independent of the magnitude
of $f$,
        $\;F^q(0)\simeq(\lambda^2\varphi^3_I/36)
        f/96\pi^2|\xi|^3>0$.
In view of the expression for $f$ the actual magnitude of this force
is again much smaller than its classical counterpart
$F^q(0)\sim|F^{\rm class}(0)|/|\xi|\ll |F^{\rm class}(0)|$. Therefore, for
the tunneling state it gives a negligible contribution to the full rolling
force. For the no-boundary state, the initial
effective force (\ref{5.37g}) vanishes, but the only effect that the
positive $F^q(0)$ can produce is that it shifts the equilibrium
point from $\varphi_I$ to slightly higher value, at which again the system
undergoes endless inflation.

At late stationary stage of inflation the dynamics of the classical background
can be approximated by the ansatz $a=\cosh[H(\varphi)t]/H(\varphi)$,
$\,\varphi\simeq\varphi_I$, and the resulting radiation
currents are dominated by the contribution of the growing mode
$\mbox{\boldmath$q$}_+\simeq\sinh Ht/3a(t)\dot\phi\to {\rm const}$, $\;Ht\gg1$:
        \begin{eqnarray}
        j_N^q=\frac{\lambda\varphi^4}4
        \frac{\lambda}{864\pi^2|\xi|^2}f, \qquad
        j_\varphi^q=\frac{\lambda\varphi^3}2
        \frac{\lambda}{864\pi^2|\xi|^2}f, \qquad Ht\gg 1.   \label{7.19}
        \end{eqnarray}
Similarly to the onset of inflation, they are strongly suppressed relative
to the classical values by the factor $\lambda/|\xi|^2\sim 10^{-10}$. The
energy density of the quantum mechanical mode,
        $\varepsilon^q=-j^q_N\simeq
        -\lambda^2|1+\delta|m_P^4/54(16\pi^2)^2|\xi|^3)$,
        $\;|\varepsilon^q|\ll m_P^4$,
is negative, while the pressure is positive $p^q=-\varepsilon^q$ \cite{efeqmy}
(anti-DeSitter case). Apparently, this is a manifestation of the ghost nature of
the
mode $\mbox{\boldmath$q$}$ whose kinetic term enters the action
(\ref{4.16}) with the wrong sign. Radiation currents (\ref{7.19}) lead to
the quantum rolling force (\ref{5.37}), $F^q\simeq 0$, which vanishes in
the leading order of the slow roll expansion.

\section{Conclusion}
\hspace{\parindent}
The dynamical contribution of the quantum mechanical mode to
effective equations turned out to be very small -- it is strongly
dominated by the effective rolling force. This property was actually
conjectured in \cite{efeq}, and now it is quantitatively confirmed.
Thus, this mode cannot change the
dynamical predictions in spatially closed model with strong non-minimal
coupling. As a model of the low-energy quantum origin of the Universe
only the tunneling state remains observationally justified, because the
no-boundary wavefunction generates infinitely long inflationary stage.
The role of this mode should not, however, be underestimated, because its
effect is model dependent, and might be important in other models generating
initial conditions for inflation. Moreover, the inflaton
mode simulates the DeSitter and Anti-DeSitter effective equations of state,
$\varepsilon+p=0$, respectively at the onset of inflation and at late times.
The sign of its energy density contribution can change depending on the
balance of the potential and kinetic terms of this ghost mode. Therefore,
it is not quite clear at the moment, what can the role of this mode be at
post inflationary epoch. A natural question arises if this mode can be
responsible for the present day observable acceleration of the Universe
as an alternative to quintessence or be capable of inducing
DeSitter-anti-DeSitter phase transitions in cosmology? This question is
subject to further studies.

\section*{Acknowledgements}
\hspace{\parindent}

This work was supported by the Russian
Foundation for Basic Research under
the grant No 99-02-16122. The work of A.O.B. was also supported by the
grant of support of leading scientific schools No 00-15-96699. The work of
D.V.N. was supported by the grant of support of leading scientific schools
No 00-15-96566.

\end{document}